\documentclass[twocolumn,amsmath,amssymb,showpacs,floatfix,prb]{revtex4}

\usepackage{graphicx}
\usepackage{dcolumn}
\usepackage{bm}
\usepackage{color}
\begin{document}

\title{Ambipolar transition voltage spectroscopy: analytical results and experimental agreement}
\author{}
\author{Ioan B\^aldea}
 \altaffiliation[Also at ]{NILPR, ISS, POB MG-23, RO 077125, Bucharest, Romania} 
\affiliation{%
Theoretische Chemie, Universit\"at Heidelberg, Im Neuenheimer Feld 229, 
D-69120 Heidelberg, Germany}

\begin{abstract}
This work emphasizes that the transition voltages $V_{t \pm}$ for both bias polarities
($V${\raisebox{1.2ex}{$_{\stackrel{>}{<}}$}$0$})
should be used to properly determine the energy offset $\varepsilon_{0}$ 
of the molecular orbital closest to electrodes' Fermi level and the bias asymmetry 
$\gamma$ in molecular junctions. Accurate analytical formulas are deduced to estimate
$\varepsilon_{0}$ and $\gamma$ solely in terms of $V_{t \pm}$. 
These estimates are validated against experiments, by showing 
that full experimental $I$-$V$-curves measured by Beebe et al [\prl {\bf 97}, 026801 (2006)] 
and Tan et al [Appl.~Phsy.~Lett.~{\bf 96}, 013110 (2010)] for both bias polarities can be excellently reproduced.
\end{abstract}


\maketitle
\section{Introduction}
\label{sec:introd}
Transition voltage ($V_t$) spectroscopy (TVS) has been recently proposed \cite{Beebe:06}
to get insight into the energy offset $\varepsilon_0$ between the metal Fermi energy and the closest 
molecular orbital (HOMO or LUMO), which has long been recognized to be a key 
quantity for the charge transport in molecular devices \cite{Datta:03}.
Due to its simplicity, it soon became very popular for interpreting molecular transport measurements.
\cite{Choi:08,Beebe:08,Chiu:08,Reed:09,Lee:09,Frisbie:10b,Choi:10,Choi:10b,Reed:10,Tan:10,Lee:11,Lennartz:11}
Within the initial proposal relying upon a tunneling barrier picture,\cite{Beebe:06}
$V_t$, the minimum of the Fowler-Nordheim (FN) plot $\ln (I/V^2)$ versus $1/V$ determined from $I$-$V$ measurements,
has been associated to the point where the barrier tilted by the applied bias change from trapezoidal 
to triangular. This yields $e V_t = \varepsilon_0$.\cite{Reed:09} 
Later, TVS was interpreted within a coherent transport model based on a single level.\cite{Huisman:09} 
In the (realistic) cases where the energy offset is sufficiently larger than the 
level broadening due to the couplings to electrodes, 
the relationship deduced within the latter model 
($e V_t = 1.15 \varepsilon_0$ for a \emph{symmetrical} orbital 
location between electrodes \cite{Baldea:2010h})
turned out to be not much different from the original ``barrier shape'' conjecture.
With certain limitations, ab initio studies \cite{Araidai:10,Thygesen:10b} give 
microscopic support to the single-level model. A significant aspect in the TVS 
analysis \cite{Thygesen:10b} is to properly account for the potential profile asymmetry. 
In the presence of this asymmetry, the FN-plots also become asymmetric,
a fact which reflects itself in different magnitudes of the transition voltages 
$V_{t +} \neq - V_{t -}$ for both bias polarities ($V${\raisebox{1.2ex}{$_{\stackrel{>}{<}}$}$0$), 
as pointed out recently.\cite{Molen:11b}
It is the main aim of this paper to consider the ambipolar TVS in detail.

The remaining part of this paper is organized as follows. The general 
theoretical framework will be presented in Sect.~\ref{sec:framework}. 
In Sect.~\ref{sec:results-tvs}, accurate analytical formulas 
will be given enabling one to directly extract the quantities of physical interest 
from the transition voltages measured for positive and negative biases. 
The accuracy and the usefulness of these theoretical formulas will be 
illustrated in Sect.~\ref{sec:applic}, 
by showing that applications to experimental $I$-$V$-data yield an excellent agreement.
Conclusions will be presented in Sect.~\ref{sec:conclusion}.
\section{Theoretical framework}
\label{sec:framework}
An applied bias $V$ affects the energy offset
$\varepsilon_0 \equiv \varepsilon_0(V)\vert_{V=0} \to \varepsilon_0(V)$. 
Most easily, as currently done in 
electrochemistry,\cite{KuznetsovElectrocheGating:00,Zhang:02,Zhang:03}
this effect can be accounted for by means of a voltage division factor \cite{Datta:03}
$\gamma$, which specifies the bias at the location of the orbital's ``center of gravity''
\begin{equation}
\label{eq-gamma}
\varepsilon_{0}(V) =  \varepsilon_{0} + \gamma \, e V .
\end{equation}
With a potential origin as in Fig.~\ref{fig:setup}, $\gamma$ ranges 
from $-1/2$ to $+1/2$.
Eq.~(\ref{eq-gamma}) corresponds to a potential that is flat
across the molecule and entirely drops at contacts.
The interfacial potential drops are 
\begin{eqnarray}
\delta V_{s} & = & V/2 - [\varepsilon_{0}(V) - \varepsilon_{0}]/e = (1/2 - \gamma) V , \nonumber \\
\delta V_{t} & = & [\varepsilon_{0}(V) - \varepsilon_{0}]/e + V/2 = (1/2 + \gamma) V .
\label{eq-delta_V_s,t}
\end{eqnarray}
A positive (negative) $\gamma$-value corresponds to a larger (smaller) 
potential drop at the positive electrode, or, alternatively, 
to a molecular orbital energy shifted upward (downward) by a positive bias.
\begin{figure}[h!]
$ $\\[6ex]
\centerline{\hspace*{-0ex}\includegraphics[width=0.4\textwidth,angle=0]{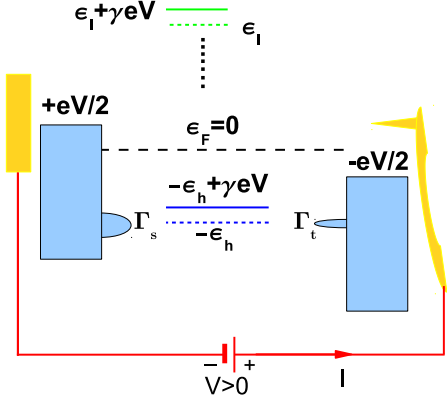}}
\caption{Schematic representation of a HOMO-mediated conduction 
(the HOMO much closer to the Fermi level than the LUMO, $\varepsilon_h \ll \varepsilon_l$) 
in a setup characterized by asymmetric 
electrodes [asymmetric voltage division $\gamma$($ > 0$) and asymmetric molecule-electrode couplings 
$\Gamma_s$ and $\Gamma_t$]. See the main text for details.}
\label{fig:setup}
\end{figure}

In the wide-band limit, wherein the transmission is Lorentzian,
the current through a single level (Newns-Anderson model) 
can be expressed analytically as 
(see, e.~g., Refs.~\onlinecite{HaugJauho,Baldea:2010e,Baldea:2010h}) 
\begin{eqnarray}
\displaystyle
I & = & N  \frac{2 e}{h}\frac{\Gamma_g^2}{\Gamma_a}
\left( \arctan \frac{\Lambda_{+}}{\Gamma_a} - \arctan \frac{\Lambda_{-}}{\Gamma_a} \right) .
\label{eq-I}
\end{eqnarray}
Here, $N$ is the (effective) number of molecules contributing 
to the current, $\Lambda_{\pm} \equiv \varepsilon_{0}(V) \pm eV/2 $.
$\Gamma_a \equiv\left( \Gamma_s + \Gamma_t\right)/2$ 
and $\Gamma_g \equiv \sqrt{\Gamma_s  \Gamma_t}$, $\Gamma_{s,t}$ being 
the level broadenings due to molecule-electrode couplings.
In usual cases of interest, $\Gamma_a \ll \vert \varepsilon_0\vert$
and voltages not much higher than $V_t$,\cite{sampled-V} the arguments of the 
inverse trigonometric functions of Eq.~(\ref{eq-I}) are large, and one can approximate
\begin{equation}
\displaystyle
I  \simeq N  \frac{2 e}{h} \Gamma_g^2 \frac{e V}{ \left( \varepsilon_0 + \gamma\, e V\right)^2 - (e V/2)^2} .
\label{eq-I-approx}
\end{equation}
\begin{figure}[h!]
$ $\\[6ex]
\centerline{\hspace*{-0ex}\includegraphics[width=0.4\textwidth,angle=0]{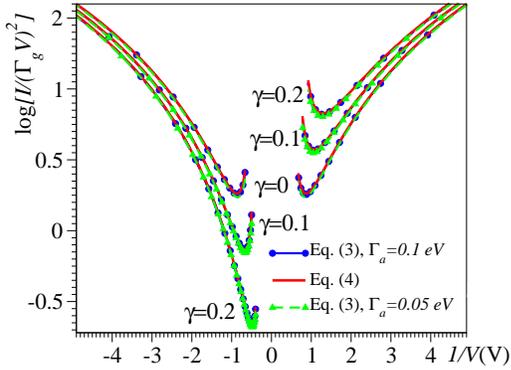}}
\caption{Fowler-Nordheim (FN) plots for a single-level coherent transport obtained 
by means of the exact Eq.~(\ref{eq-I}) and approximate Eq.~(\ref{eq-I-approx})
for symmetric ($\gamma = 0$) and asymmetric ($\gamma \neq 0$) bias profiles.
They show that for level broadenings $\Gamma_a$ sufficiently smaller than 
its energy offset ($\varepsilon_0 = 1$\,eV), the FN-transition is accurately 
described by  Eq.~(\ref{eq-I-approx}). 
See the main text for details.}
\label{fig:FN_ex_vs_asympt}
\end{figure}

The comparison with the results based on the exact Eq.~(\ref{eq-I}) 
shows that Eq.~(\ref{eq-I-approx}) very accurately describes the 
FN-transition; see Fig.~\ref{fig:FN_ex_vs_asympt}.
\section{Useful analytical results for transition voltage spectroscopy}
\label{sec:results-tvs}
Eq.~(\ref{eq-I-approx}) can be used to deduce
simple analytical expressions of the relevant quantities 
within the (realistic) assumption of 
a level broadening sufficiently smaller than the energy offset,
which are exact to $  \mathcal{O}\left(\Gamma_{a}/\varepsilon_{0}\right)^2$.
By imposing $\partial \ln (I/V^2) / \partial (1/V) =0 $, one obtains 
the transition voltages $V_{t 1,2}$ as
\begin{eqnarray}
\chi_{t 1} & \equiv & \varepsilon_{0}/V_{t 1} = - 2 \gamma + \sqrt{\gamma^2 + 3/4} , \nonumber \\
\chi_{t 2} & \equiv & \varepsilon_{0}/V_{t 2} = - 2 \gamma - \sqrt{\gamma^2 + 3/4} .
\label{eq-ut_2}
\end{eqnarray}
Because $-1/2 \leq \gamma \leq 1/2$, it is easy to show that $\chi_{t 1} > 0$ and $\chi_{t 2} < 0$. 
Therefore, the signs of $V_{t 1}$ and $V_{t 2}$ are opposite.
Denoting by $V_{t +}$($> 0$) 
and $V_{t -}$($< 0$) the transition voltage for positive and negative polarities,
$V_{t +} \equiv V_{t 1}$ 
and $V_{t -} \equiv V_{t 2}$ for LUMO-mediated transport ($\varepsilon_0 > 0$),
while for HOMO-mediated transport ($\varepsilon_0 \equiv - \varepsilon_{h} < 0$)
$V_{t +} \equiv V_{t 2}$ and $V_{t -} \equiv V_{t 1}$. In the HOMO case,
$V_{t +} < \vert V_{t -}\vert$ for $\gamma > 0$, whereas $V_{t +} > \vert V_{t -}\vert$ for $\gamma < 0$.
For $\gamma = 0$, the result $\vert V_{t\pm}/\varepsilon_{0}\vert = 2/\sqrt{3} = 1.15$
for symmetric case \cite{Baldea:2010h} is recovered.

Concerning the signs in general, it is worth noting that, according to Eq.~(\ref{eq-gamma}), 
a redefinition of the bias 
polarity ($V \to -V$) yields a sign change in the division potential factor ($\gamma \to -\gamma$).
Therefore, the discussion can be restricted to the range, e.~g., $0 \leq \gamma \leq 1/2$. 

To illustrate the accuracy of the transition voltages $V_{t\pm}$ 
expressed by the above analytical formulas,
a comparison with the transition voltages deduced from the exact Eq.~(\ref{eq-I}) 
is presented in Fig.~\ref{fig:Vt_ex_vs_asympt}. 
\begin{figure}[h!]
$ $\\[6ex]
\centerline{\hspace*{-0ex}\includegraphics[width=0.4\textwidth,angle=0]{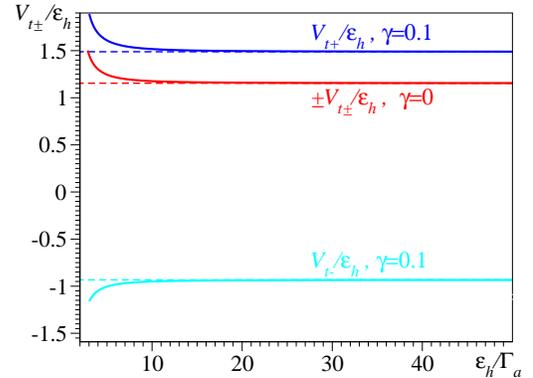}}
\caption{The reduced transition voltages $V_{t \pm}/\varepsilon_{h}$ 
plotted versus the inverse of the relative HOMO-level ($\varepsilon_{0} = -\varepsilon_{h}$) 
broadening $\Gamma_{a}/\varepsilon_h$ 
for given potential division factors ($\gamma=0, 0.1$) deduced exactly 
[Eq.~(\ref{eq-I}), thick lines] and approximately [Eq.~(\ref{eq-I-approx}), thin horizontal lines].
Notice the rapid saturation 
for sufficiently small level broadenings $\Gamma_{a}$.
See the main text for details.}
\label{fig:Vt_ex_vs_asympt}
\end{figure}
As visible there, 
the exact results are rapidly approached for small $\Gamma_{a}/\varepsilon_{0} \to 0$.
Graphical results obtained by means of Eq.~(\ref{eq-ut_2}) are presented in 
Fig.~\ref{fig:ut1_ut2} for values $0 \leq \gamma \leq 1/2$. 
In the opposite case ( $-1/2 \leq \gamma < 0$) they can be deduced by symmetry
$V_{t \pm}(-\gamma) = \pm V_{t \mp}(\gamma)$.

An important consequence of Eq.~(\ref{eq-ut_2}) is that both 
the voltage division factor $\gamma$ and the energy offset 
$\varepsilon_{0}$ can be determined from $V_{t \pm}$ \cite{difference-molen}
\begin{eqnarray}
\vert \varepsilon_{0}\vert & = & 2 \frac{ e \, \vert V_{t +} V_{t -}\vert}
           {\sqrt{V_{t +}^2 + 10 \vert V_{t +} V_{t -}\vert / 3  + V_{t -}^2}} ,
\label{eq-e0_2} \\
\gamma & = &  \frac{\mbox{sign\,} \varepsilon_{0}}{2} 
\frac{V_{t +} + V_{t -}}
     {\sqrt{V_{t + }^2 + 10  \vert V_{t +} V_{t -}\vert /3 + V_{t -}^2}} .
\label{eq-gamma_2} 
\end{eqnarray}
\begin{figure}[h!]
$ $\\[6ex]
\centerline{\hspace*{-0ex}\includegraphics[width=0.4\textwidth,angle=0]{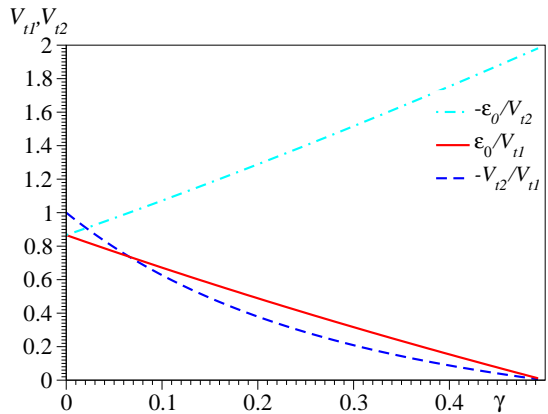}}
$ $\\[6ex]
\centerline{\hspace*{-0ex}\includegraphics[width=0.4\textwidth,angle=0]{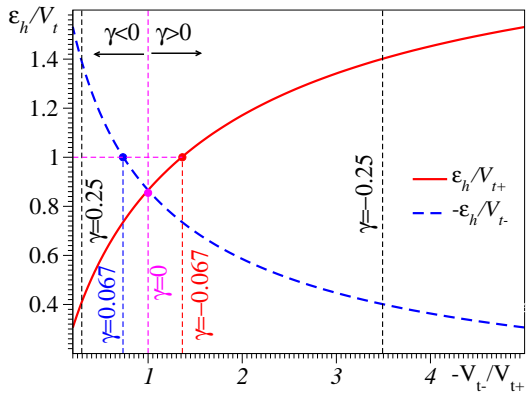}}
\caption{(a) Results for the transition voltages $V_{t \pm}$ 
as a function of potential division factor $\gamma$
deduced from Eq.~(\ref{eq-ut_2}).
(b) Ratio between the energy offset $\varepsilon_{0} = - \varepsilon_{h} < 0$ 
and the transition voltages $V_{t \pm}$ 
of both bias polarities as a function of the transition voltage asymmetry 
$V_{t -} / V_{t +}$. Each point of this curve corresponds to a given bias asymmetry 
$\gamma$, and a few values important for the discussion in the main text are indicated:  
$\gamma = 0.25 \to \vert V_{t,max} / V_{t,min}\vert = 3.5$, 
$\gamma = 0.067 \to  \vert \varepsilon_{0}/V_{t, max}\vert = 0.40,  
\vert V_{t-}\vert = \vert \varepsilon_{0}\vert$, 
and $\vert V_{t, min} / V_{t,min}\vert = 1.36$.
 See the main text for details.}
\label{fig:ut1_ut2}
\end{figure}

TVS's proof of value for molecular electronics is the fact that the FN-minimum occurs
at voltages below the values corresponding to resonant tunneling
where the differential conductance is maximum. 
This is important
because, with seldom exceptions,\cite{Wandlowski:08,Lennartz:11} 
molecular junctions cannot withstand such high voltages. 
Still, as already noted,\cite{Beebe:08,Baldea:2010h}
it is only a small range $V > V_t$ (if at all \cite{Reed:04}) that 
can be sampled in experiments. The situation can be 
further improved by using the minimum $V = V_{t \pm}(\kappa)$
of a generalized FN-plot $\ln (I/V^\kappa)$ vs.~$1/V$ 
($1 < \kappa \leq 2$).\cite{Thygesen:11}
General analytical expressions valid for arbitrary $\kappa$ can also be deduced 
\begin{equation}
\chi_{t1,2}({\kappa}) \equiv \frac{\varepsilon_{0}}{V_{t1,2}(\kappa)} = \frac{1}{\kappa - 1}
\left( - \kappa \gamma \pm \sqrt{\gamma^2 + \frac{\kappa^2 - 1}{4}}\right) ,
\label{eq-ut_kappa} 
\end{equation}
\begin{equation}
\vert \varepsilon_{0}\vert = \frac{\kappa(\kappa + 1)}{\kappa^2 - 1} 
\frac{e\, \vert V_{t +}(\kappa)  V_{t -}(\kappa)\vert}
     {\sqrt{V_{t + }^2(\kappa) + 
         2 \frac{\kappa^2 + 1}{\kappa^2 - 1} \vert V_{t +}(\kappa) V_{t -}(\kappa)\vert 
         + V_{t -}^2(\kappa)}} ,
\label{eq-e0_kappa}
\end{equation}
\begin{equation}
\gamma = \frac{\mbox{sign\,}\varepsilon_{0}}{2} 
\frac{V_{t +}(\kappa) + V_{t -}(\kappa)}
     {\sqrt{V_{t + }^2(\kappa) + 
         2 \frac{\kappa^2 + 1}{\kappa^2 - 1} \vert V_{t +}(\kappa) V_{t -}(\kappa)\vert 
         + V_{t -}^2(\kappa)}} .
\label{eq-gamma_kappa}
\end{equation}
The assignment $ V_{t 1,2}(\kappa) \to  V_{t \pm}(\kappa)$ is the same as discussed above
for $\kappa = 2$.
The usage of $\kappa$-values smaller than 2 
results in lower transition voltages $ V_{t \pm}(\kappa)$,\cite{Thygesen:11} 
which can easier be sampled experimentally. On the other side, being smaller they are
more affected by the relatively large ($ > 0.1$\,V \cite{Tan:10,Lennartz:11}) 
experimental errors.

From a pragmatic standpoint, Eqs.~(\ref{eq-e0_2}) and (\ref{eq-gamma_2})
as well as the more general Eqs.~(\ref{eq-e0_kappa}) and (\ref{eq-gamma_kappa})
represent the core of this paper: they enable one to determine the quantities 
of physical interest $\varepsilon_{0}$ and $\gamma$ from the transition voltages 
$V_{t \pm}$ measured for both bias polarities.
\section{Discussion of experimental data}
\label{sec:applic}
The above analytical results hold in general for
the transport mediated by a
single level whose energy offset is sufficiently larger than its electrode-induced 
broadening.

I will next apply these results to the experimental data of 
Refs.~\onlinecite{Beebe:06} and  \onlinecite{Tan:10},
obtained for the HOMO-mediated transport through molecular junctions in CP-AFM 
(conducting probe-atomic force microscopy) 
\cite{Beebe:06,Tan:10} and CW (crossed-wire) \cite{Beebe:06} setups.
For a concrete comparison with experiment, it is obviously important 
to correctly assign, out of
the two transition voltages measured for opposite bias polarities
in experiment, which is $V_{t +}$ and which is $V_{t -}$.  
Therefore, before entering into details, I note that the positive and negative biases 
have been chosen as the ones utilized in the experiments of Refs.~\onlinecite{Beebe:06,Tan:10}
discussed below.

Table \ref{table} collects the experimental data for $V_{t \pm}$ from Refs.~\onlinecite{Beebe:06} and  \onlinecite{Tan:10}.
They have been used to compute the  
values of $\varepsilon_{h}$ and $\gamma$ from Eqs.~(\ref{eq-e0_2}) and (\ref{eq-gamma_2}),
which are also given in Table \ref{table}.

Further, I will use these values of $\varepsilon_{h}$ and $\gamma$ 
to reproduce 
the available $I$-$V$-data for anthracenethiol- and terphenylthiol-based junctions
measured in Refs.~\onlinecite{Beebe:06} 
and \onlinecite{Tan:10}\cite{thanks-pramod}), respectively.
To this aim, I will employ 
Eq.~(\ref{eq-I-approx}), which represents a very good approximation of the exact 
Eq.~(\ref{eq-I}) for biases not too much larger than the transition voltages.
The prefactor $N \Gamma_{g}^2$ in Eq.~(\ref{eq-I-approx}) can be determined 
from the experimental linear conductance
($N^{1/2}\Gamma_{g} = 0.051$\,eV and $N^{1/2}\Gamma_{g} = 0.124$\,eV, respectively).
The theoretical curves obtained in this manner are plotted
against the experimental ones in Figs.~\ref{fig:iv_BeebeTan}a and b.
The agreement is excellent, and this demonstrates the remarkable accuracy of the present approach
based on Eqs.~(\ref{eq-I-approx}) and (\ref{eq-gamma}).\cite{individual-Gammas} 

Table \ref{table} shows that for the molecular junctions 
of Refs.~\onlinecite{Beebe:06} and \onlinecite{Tan:10} 
the energy offset is, in fact, very close to the 
estimate of the barrier-shape conjecture if the transition voltage for 
positive biases is employed, $e V_{t +} = \varepsilon_{h}$.\cite{Reed:09} 
As visible in Fig.~\ref{fig:ut1_ut2}b, this does not hold in general but only for potential division 
factors close to $\gamma = 0.067$.
\begin{table}[h!]
\begin{center}
\begin{tabular}{l@{\hspace{0ex}}c@{\hspace{1ex}}c@{\hspace{1ex}}c@{\hspace{1ex}}c@{\hspace{0ex}}c}
\hline
\hline
Molecule, platform & $V_{t +}$\,(V) & $V_{t -}$\,(V) & $\varepsilon_{h}$\,(eV) & $\gamma$ \\
\hline
Anth-SH, CP-AFM $^a$ & 0.62 & -0.85 & 0.62 & 0.068 \\
Anth-SH, CW $^a$ & 0.57 & -0.85 & 0.59 & 0.086 \\
TP-SH, CP-AFM $^a$ & 0.67 & -0.82 & 0.64 & 0.044 \\
TP-SH, CP-AFM $^b$ & 0.69 & -0.85 & 0.69 & 0.069 \\
TP-SH, CW $^a$ & 0.66 & -0.92 & 0.67 & 0.071 \\
\hline
\end{tabular}
\end{center}
\caption{Experimental values of $V_{t\pm}$ for anth(racene)- and terphenyl (TP)-based 
molecules and platforms  
from Refs.~\onlinecite{Beebe:06} ($a$) and \onlinecite{Tan:10} ($b$), and 
values of $\varepsilon_{h}$ and $\gamma$ calculated via Eqs.~(\ref{eq-e0_2})
and (\ref{eq-gamma_2}). See the main text for details.}
\label{table}
\end{table}

\begin{figure}[h!]
$ $\\[6ex]
\centerline{\hspace*{-0ex}\includegraphics[width=0.4\textwidth,angle=0]{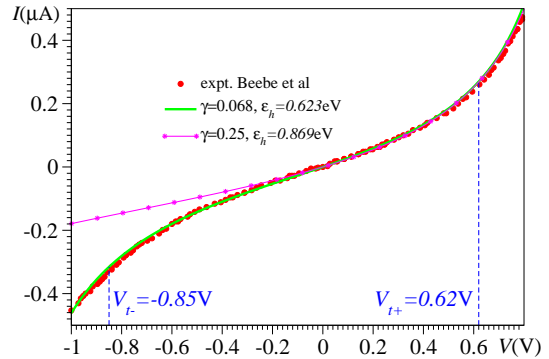}}
$ $\\[6ex]
\centerline{\hspace*{-0ex}\includegraphics[width=0.4\textwidth,angle=0]{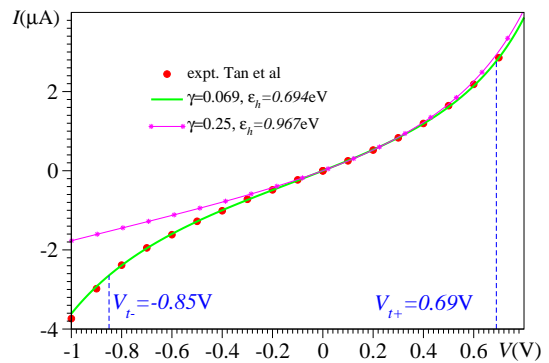}}
\caption{The experimental $I$-$V$-curves measured in CP-AFM setup by 
Beebe et al \cite{Beebe:06} and Tan et al \cite{Tan:10,thanks-pramod} 
for anthracene- and terphenyl-based junctions, respectively 
plotted along with the theoretical curves computed via Eq.~(\ref{eq-I-approx}) 
with the $\varepsilon_{h}$- and $\gamma$-values of Table \ref{table}.
Also shown are the theoretical curves deduced by using
the value $\gamma = 0.25$ given in Ref.~\onlinecite{Thygesen:10b}, which completely
disagree with experiments for negative voltages. 
See the main text for details.}
\label{fig:iv_BeebeTan}
\end{figure}

Noteworthy, all the cases presented in Table \ref{table} are characterized by 
small positive $\gamma$-values, revealing that the potential drop 
at the soft contact (e.~g., AFM-tip) is slightly larger, $\delta V_{t} \agt \delta V_{s}$
[cf.~Eq.~(\ref{eq-delta_V_s,t})].
So, even a (very) small difference in the interfacial 
potential drops causes a significant polarity dependence of the 
transition voltage.
This suggests that, more than the active molecule (which can differ, see Table \ref{table}),
the contacts are important for the $V_{t \pm}$-asymmetry 
and calls for a systematic experimental investigation on the role of the 
contact groups (thiol, amine, etc).

The $\gamma$-values of Table \ref{table}
are significantly smaller than those estimated via DFT calculations 
($\gamma \simeq 0.2 - 0.3$)\cite{Thygesen:10b} and that of $\gamma = 0.25$
claimed \cite{Thygesen:11} to be appropriate 
for the experiments of Refs.~\onlinecite{Beebe:06} and \onlinecite{Tan:10}.
With the value $\gamma = 0.25$, Eq.~(\ref{eq-ut_2}) yields HOMO-offsets $\varepsilon_{h} = 0.87$\,eV 
(for $V_{t +} = 0.62$\,V) and $\varepsilon_{h} = 0.97$\,eV (for $V_{t +} = 0.69$\,V)
for the $I$-$V$-curves shown in Refs.~\onlinecite{Beebe:06} and \onlinecite{Tan:10},
respectively. 
The very small difference (much smaller than the experimental inaccuracies) 
between the above values and those 
given in Ref.~\onlinecite{Thygesen:11} ($\varepsilon_{h} \approx 0.85$\,eV and $1.0$\,eV),
deduced by assuming a Lorentzian transmission 
[the key assumption underlying Eq.~(\ref{eq-I})\cite{Baldea:2010e}],
demonstrates again the high accuracy of the present estimates based on Eq.~(\ref{eq-I-approx}).

While the $\varepsilon_{h}$-estimates 
($\varepsilon_{h} = 0.62$\,eV and $\varepsilon_{h} = 0.69$\,eV deduced in the present paper versus
$\varepsilon_{h} = 0.85$\,eV and $\varepsilon_{h} = 1.0$\,eV of Ref.~\onlinecite{Thygesen:11})
cannot be directly checked against experiment,
there is a conclusive test to decide which potential division factor 
($\gamma \simeq 0.07$ or $\gamma = 0.25$) is correct. 
Namely, for $\gamma = 0.25$, the transition voltages for
negative bias are $V_{t -} = -2.17$\,V and $V_{t -} = -2.42$\,V, and they clearly disagree with 
the experimental data of  Refs.~\onlinecite{Beebe:06} and \onlinecite{Tan:10}, respectively.

It is worth mentioning that, as visible in Fig.~\ref{fig:ut1_ut2},
the $V_{t +}$/$V_{t -}$ asymmetry very rapidly varies with the division potential factor $\gamma$. It becomes 
important even for small $\gamma$, as revealed by Table \ref{table}.
Although more or less asymmetric $I$-$V$-characteristics 
[$I(V) \neq - I(-V)$] are ubiquitous in 
molecular electronics, such a large asymmetry (a factor $\simeq 3.5$) 
between the positive and negative transition
voltages  ($V_{t +} = 0.62$\,V and $V_{t +} = 0.69$\,V versus
$V_{t -} = -2.17$\,V and $V_{t -} = -2.42$\,V, respectively) has not been reported so far.

A curious aspect should still be noted at this point. 
By using the values $\varepsilon_{h} = 0.87$\,eV and $\varepsilon_{h} = 0.97$\,eV
(the curves are indistinguishable from the choice $\varepsilon_{h} = 0.85$\,eV and $\varepsilon_{h} = 1.0$\,eV),
and $\gamma = 0.25$, and by adjusting the prefactor in Eq.~(\ref{eq-I-approx}) to 
reproduce the experimental linear conductance, the agreement with experiments 
for \emph{positive} biases is also good; see Figs.~\ref{fig:iv_BeebeTan}a and b.
However, as visible there, there is a complete disagreement for negative biases. 

Although the DFT inability to solve level alignment problems is well 
known,\cite{Neaton:06,Thygesen:09c} it has been claimed \cite{Thygesen:10b,Thygesen:11}
that DFT-estimates of the ratio $\varepsilon_{0}/V_t$ could be trusted. The
DFT-values of $\gamma$($\sim 0.2 - 0.3$) are based just on the fact that 
the ratio $\varepsilon_{0}/V_t$ is determined by the value of $\gamma$.
The above analysis reveals a further limitation of the DFT-based approaches to molecular transport,  
raising serious doubts on its reliability for TVS studies.

One should finally note that the above quantity 
$\varepsilon_{h}$ is to be understood
as the HOMO-offset characterizing a molecule linked to \emph{two} (non-biased) electrodes.
Therefore, it is not surprising that the present values 
($\varepsilon_{h} = 0.62$\,eV and $\varepsilon_{h} = 0.69$\,eV  
for anthracenthiol- and terphenylthiol-based junctions)
deduced from transport data can differ from the ultraviolet photoelectron spectroscopy (UPS) data.\cite{Beebe:06} 
Indeed, they do differ: 
UPS measurements on anthracene- and terphenyl-thiol on gold yielded 
$\varepsilon_{h}^{UPS} \simeq 1.7$\,eV and $1.8$\,eV.\cite{Beebe:06} Partly,
the difference between $\varepsilon_{h}$ and $\varepsilon_{h}^{UPS}$ can be attributed 
to image effects,\cite{Schmickler:95,Thygesen:11} but I do not address this issue here.
\section{Conclusion}
\label{sec:conclusion}
Although the role of the asymmetric interfacial potential drops has already 
been noted in the original TVS paper \cite{Beebe:06} and considered 
more recently,\cite{Thygesen:10b,Thygesen:11,Molen:11b}
a direct quantitative analysis of full $I$-$V$-curves measured in experiments 
has not been attempted in previous studies.  
The present paper shows that, 
by resorting to ambipolar TVS, it is possible to determine not only the energy offset
$\varepsilon_{0}$ (at which the original TVS aimed), but also the potential 
division factor  $\gamma$ without a notably increased effort: 
the only quantities to be determined experimentally are the transition voltages 
$V_{t \pm}$ for both bias polarities. For quantitative estimates, simple, very
accurate analytical 
formulas have been deduced, which have been excellently 
validated against available experimental data. 
The excellent agreement found in the present paper gives a strong support 
to the correctness of the Lorentzian transmission [the key assumption underlying Eqs.~(\ref{eq-I})
and (\ref{eq-I-approx})] and rules out possible higher-order contributions ($\sim V^2, V^3$, etc)
to the RHS of Eq.~(\ref{eq-gamma}) for the molecular junctions investigated in Refs.~\onlinecite{Beebe:06}
and \onlinecite{Tan:10}. 

The present analysis has emphasized the need to include \emph{both} bias polarities 
in order to obtain correct $\varepsilon_{0}$- and $\gamma$-estimates. 
In this context, a specification for the discussion \cite{Reed:11,Lee:11}
of experimental data is helpful.
According to a result of previous work,\cite{Thygesen:10b}
the ratio between the energy offset and the transition voltage 
$\vert \varepsilon_{0}\vert /V_{t}$ varies from 0.86 to 2. The present paper
reconfirms this result with an important specification: the aforementioned range
refers to ratio $\vert \varepsilon_{0}\vert /V_{t, min}$, which corresponds to 
the transition voltage of the \emph{smallest} magnitude
$V_{t, min} \equiv \min\left( V_{t +}, - V_{t -}\right)$. 
The ratio $\vert \varepsilon_{0}\vert /V_{t, max}$, which corresponds to 
the transition voltage of the \emph{largest} magnitude
$V_{t, max} \equiv \max\left( V_{t +}, - V_{t -}\right)$, ranges from $0$ to $0.866$
[cf.~Eq.~(\ref{eq-ut_2}) and Fig.~\ref{fig:ut1_ut2}a].
\section*{Acknowledgments}
The author thank Horst K\"oppel for valuable discussions and to Pramod Reddy 
for providing him the raw experimental data used in Fig.~\ref{fig:iv_BeebeTan}b. 
Financial support for this work provided by the Deu\-tsche 
For\-schungs\-ge\-mein\-schaft (DFG) is gratefully acknowledged.

\begin{thebibliography}{10}
\bibitem{difference-molen}
Eq.~(\ref{eq-ut_2}) yields expressions different from those given without
  deduction in Ref.~\onlinecite{Molen:11b}. For LUMO-mediated transport,
  ($\varepsilon_{0} > 0$), Ref.~\onlinecite{Molen:11b} gives $ \vert V_{t
  -}\vert /V_{t +} = (1 - 2 \gamma)/(1 + 2 \gamma)$, which differs from the
  result $ \vert V_{t -}\vert /V_{t +} = (\sqrt{3/4 + \gamma^2} - 2
  \gamma)/(\sqrt{3/4 + \gamma^2} + 2 \gamma)$ derived from Eq.~(\ref{eq-ut_2}).
  Notice that the quantity denoted by $\eta$ in Ref.~\onlinecite{Molen:11b} is
  related to the presently used $\gamma$ 
  (in the standard notation of electrochemistry \cite{KuznetsovElectrocheGating:00,Zhang:02,Zhang:03}) by $\eta = 1/2 + \gamma$.
\bibitem{individual-Gammas}
The present method permits an upper bound ($N \geq 1$) estimate of the
  geometrical average $\Gamma_{g} = \sqrt{\Gamma_{s} \Gamma_{t}}$ but not of
  the individual electrodes' hybridizations $\Gamma_{s,t}$; presumably, the
  coupling to the soft contact is smaller $\Gamma_{t} <
  \Gamma_{s}$. The arithmetic average $\Gamma_{a} = (\Gamma_{s} +
  \Gamma_{t})/2$ could be deduced by using the exact Eq.~(\ref{eq-I}) instead
  of Eq.~(\ref{eq-I-approx}), but the large experimental inaccuracies ($\delta
  V_{t} > 0.1$\,V \cite{Beebe:06,Tan:10}) preclude reliable evaluations.
\end{thebibliography}

\end{document}